\documentclass[letterpaper, 10 pt, conference]{ieeeconf}  

\IEEEoverridecommandlockouts                              

\usepackage{graphics,subfig} 
\usepackage{epsfig} 
\usepackage{amsmath,color} 
\usepackage{amssymb}  
\usepackage{booktabs}
\usepackage[ruled,vlined]{algorithm2e}

\def\qedsymbol{\ensuremath{\Box}}      

\def\qed{\ifhmode\unskip\nobreak\fi\quad\qedsymbol}     
\def\frqed{\ifhmode\nobreak\hbox to5pt{\hfil}\nobreak%
\hskip 0pt plus1fill\nobreak\fi\quad\qedsymbol\renewcommand{\qed}{}} 

\def\QEDsymbol{\vrule width.6em height.5em depth.1em\relax}
\def\frQED{\ifhmode\nobreak\hbox to5pt{\hfil}\nobreak%
\hskip 0pt plus1fill\nobreak\fi\quad\QEDsymbol\renewcommand{\qed}{}} 
\def\QED{\ifhmode\unskip\nobreak\fi\quad\QEDsymbol}     

\newtheorem{theorem}{Theorem}[section]
\newtheorem{proposition}[theorem]{Proposition}
\newtheorem{corollary}[theorem]{Corollary}

\DeclareMathOperator*{\argmax}{arg\,max}

\newcommand{\Val}{\operatorname{Val}}
\renewcommand{\det}[1]{\operatorname{det}(#1)}
\newcommand{\R}{\mathbb{R}}
\newcommand{\abs}[1]{|#1|}

\renewcommand{\baselinestretch}{0.97}

\title{\LARGE \bf
Secure Route Planning Using Dynamic Games with Stopping States
\author{Sandeep Banik \qquad Shaunak D. Bopardikar
\thanks{The authors are with the Department of Electrical and Computer Engineering at Michigan State University, East Lansing, MI, USA. Emails: \texttt{baniksan@msu.edu; shaunak@egr.msu.edu}}
}
}

\begin{document}

\maketitle
\thispagestyle{empty}
\pagestyle{empty}

\begin{abstract}
We consider the classic motion planning problem defined over a roadmap in which a vehicle seeks to find an optimal path from a source to a destination in presence of an attacker who can launch attacks on the vehicle over any edge of the roadmap. The vehicle (defender) has the capability to switch on/off a countermeasure that can detect and permanently disable the attack if it occurs concurrently. We model the problem of traveling along an edge using the framework of a simultaneous zero-sum dynamic game (edge-game) with a stopping state played between an attacker and defender. We characterize the Nash equilibria of an edge-game and provide closed form expressions for two actions per player. We further provide an analytic and approximate expression on the value of an edge-game and characterize conditions under which it grows sub-linearly with the number of stages. We study the sensitivity of Nash equilibrium to the (i) cost of using the countermeasure, (ii) cost of motion and (iii) benefit of disabling the attack. The solution of an edge-game is used to formulate and solve for the secure planning problem known as a meta-game.  We design an efficient heuristic by converting the problem to a shortest path problem using the edge cost as the solution of corresponding edge-games. We illustrate our findings through several insightful simulations.
\end{abstract}

\section{Introduction}
The rise of connected autonomous vehicles (CAV) and unmanned aerial vehicles (UAV) have paved its path into various applications including mobility services, vehicle platooning, delivery systems, search and rescue operations and surveillance. However, the use of such vehicles faces a number of challenges including cyber and physical attacks~\cite{UAV-cyber-attack,UAV-sec_vulnerabilities,UAV-cyber-sec-ground-control}. Continuous scanning for attacks requires energy which can severely hamper the successful completion of the mission. Therefore, it becomes imperative to strike a balance between scanning and conserving energy at the route planning stage, which motivates research in the domain of \emph{secure route planning}. This work will provide a significant impact in assuring integrity of various real-world systems such as Amazon Prime Air or Google's Project Wing. This paper studies the interplay between costs related to mobility and security in a prototypical path planning problem in presence of an adversary via a dynamic game with stopping states.

\subsection{Related literature}
Security of cyber-physical systems (CPS) including mobile robots has attracted a lot of attention recently. Recent works include~\cite{pajic2014robustness,lee2015secure} that characterizes robustness of state estimators under attack in presence of process noise and modeling errors. Reference~\cite{hespanha2019output} addresses the use of game theoretic methods to compute locally optimal solutions in presence of attacks and known bounds on disturbances applied to GPS-spoofing of a linear dynamical system. Reference~\cite{liu2020secure} determines a secure trajectory for a robot (autonomous vehicle) when moving a source to a destination and characterize conditions under which attack remains undetected. 

\medskip

The importance of data communication in order to avoid damages or manipulate systems is discussed in \cite{Agarwal2018DistortingAA}. The authors develop a metric to secure CPS communication and mitigate potential attacks. Owing to the increase in swarm-robotics applications, reference \cite{DisAtt-MultiA-Pratap2019}
proposes a distributed robust sub-modular optimization algorithm that will safeguard robots against denial-of-service failures/attacks. Appending to swarm-robotics, reference~\cite{Tsiamis2019MotionPW} addresses security measures for motion planning to safeguard a mobile robot against eavesdropping. The proposed solution is a secure communication in order to encode these packets containing trajectory information to save it from any eavesdropper. Reference \cite{Krause2011RandomizedSI}
addresses the problem of designing a sensor network to optimally secure critical infrastructure, such as a helicopter deployed in a flood hit region to search and relocate the survivors and efficiently obtain near-optimal distributions. 

\medskip

Game theory can be used to model cyber-security aspects of vehicular networks such as communication links, hardware and software of such systems~\cite{alpcan2010network}. Attacks over the wireless communication channels have been studied in~\cite{UAV-cyber-attack}. Cyber-security threats are modeled and analyzed to generate the threat profile of a system in order to secure various  vulnerabilities. Reference~\cite{UAV-cyber-sec-ground-control} models cyber-security in conjunction with the communication links, hardware and software to develop a risk model of a threat profile. A Man-in-the-Middle attack was demonstrated in reference~\cite{UAV-sec_vulnerabilities}. It was shown that professional UAV(s) used for mission critical tasks are susceptible to such attacks and can be secured through an appropriate set of countermeasures. 

\medskip

In the context of secure route planning, recent works to the best of our knowledge are considered in~\cite{Sanjab2017ProspectTF, Sanjab2019GameofDrones}, which adopt a game theoretic model to select a route for a UAV in presence of a malicious attacker that can attack on vertices of the domain. Our work differs from these references in multiple aspects including: dynamic modeling of the attack process over an edge as opposed to attacks being static and on vertices; and no specific requirements on the edges over multiple paths to be non-overlapping.

\subsection{Contribution}
\begin{figure}[h]
	\centering
	\includegraphics[width=.48\textwidth]{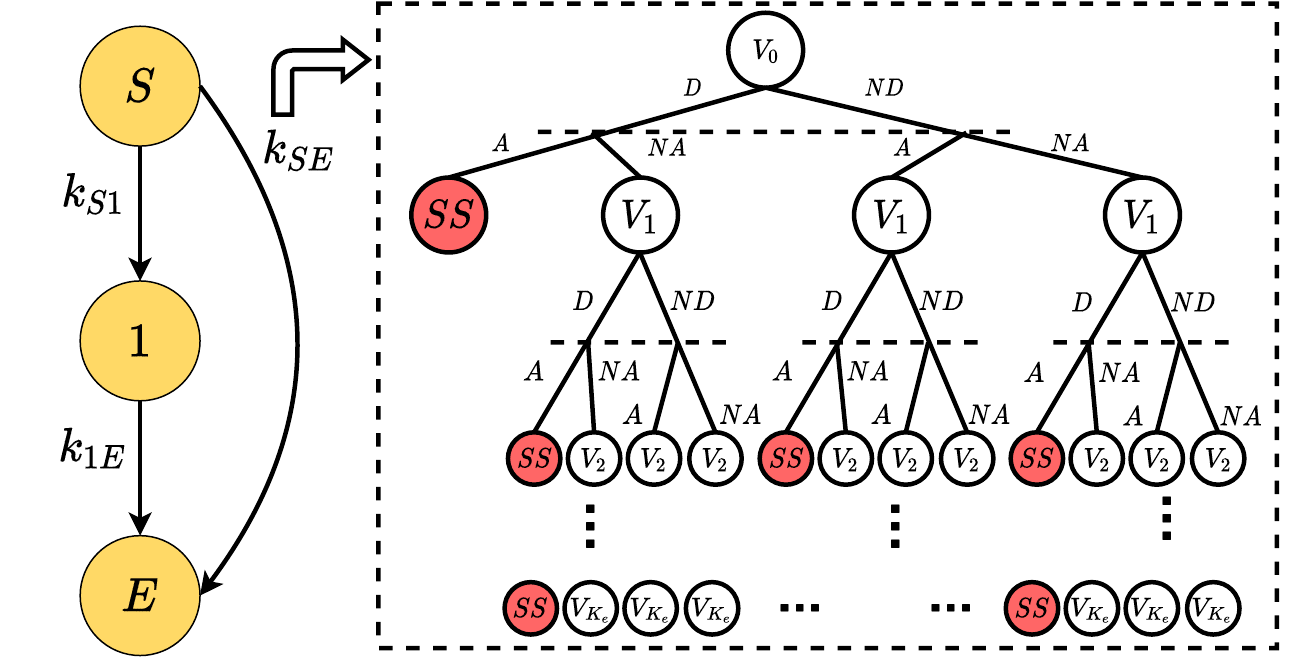}
	\caption{\small An edge-game along the edge $SE$. The dotted line at each stage indicate the choice of attacker. The policy of the attacker (resp. defender) is abbreviated as $\{A,NA\}$ (resp. $\{D,ND\}$) for $\{\text{Attack, No attack}\}$ (resp. $\{\text{Defend, No defend}\}$). The termination of the game is indicated by the stopping state $SS$.}
	\label{fig:MZSG-illustration}	
	\vspace{-0.18in}
\end{figure}

The contributions of this work are three-fold. First, we introduce a framework for multistage zero-sum games with a novel structure of a \emph{stopping state}, i.e., the game terminates if the players play out of a given subset of their actions (cf.~Figure~\ref{fig:MZSG-illustration}). This game serves as a model to characterize the costs of mobility and security while moving over an edge of a given roadmap. The proposed model can be considered as a dual version of the classic \emph{Chicken game} or the \emph{War of attrition} \cite{hespanha2017noncooperative} with an additive stage cost which models look-ahead. We characterize the Nash equilibria of this game and completely analyze the case of two actions per player. 

Second, we construct a meta-game over the roadmap, where the objective of the defender (UAV or CAV) is to go from a source node to a destination node while minimizing the impact of attack. The attacker's objective is to target the most vulnerable edge over the given set of paths over which the defender plans its path(s). Since both the attacker and defender are energy constrained devices, we constraint both to defend and attack over an edge of a roadmap. Over a simple network, we quantify the sensitivity of a path (resp. edges for the attacker) to the costs of mobility and security.

Third, for larger roadmaps, the meta-game solution is compared against a simple heuristic based on computing the shortest path constrained to an edge defense and attack for a range of graph sizes. The shortest path is calculated by replacing the edge weights with value of playing the edge-game along the corresponding edges. We observe that in sparse graphs, the meta-game approach yields the same solution as the heuristic with a high likelihood, whereas it yields a reduced path length in dense graphs, but with a larger computation time.  

\subsection{Outline of the paper}
The paper is organized as follows. We formulate the edge-game and meta-game in  Section~\ref{sec:model}. In Section~\ref{sec:mult-stage-game-edge}, the edge-game solution and its dependency on mobility and security cost is determined. The sensitivity of edge-game is studied with the game parameters. The meta-game is evaluated in Section~\ref{sec:meta-game} on a simple graph and the sensitivity of choosing the shortest path versus alternate ones are evaluated with respect to the game parameters. In Section~\ref{sec:simulation}, the meta-game is simulated over large graphs and compared against a shortest path heuristic. Finally, we conclude this paper and identify future directions in Section~\ref{sec:conclusion}. The proofs of all mathematical claims are presented in the appendix.

\section{Problem Formulation}
\label{sec:model}
Consider an environment modeled as a graph $G$ with a set of vertices $V$ and a set of edges $E$. We model the travel from a vertex to another over an edge e, as a multi-stage zero-sum finite game known as an edge-game (e-game) defined by a sequence of matrices $S^1, S^2, \dots, S^{K_e}$, where $K_e$ is the number of stages for the edge $e$. For ease of exposition, we restrict the description to the case where each matrix $S^k \in \mathbb{R}^{2\times 2}$. For a given edge $e$, at any stage k, when the defender(row player) and attacker (column player) simultaneously select action from the set \{Defend, No Defend\} and \{Attack, No Attack\} respectively, the attacker payoff equals $S^k_{ij}$. We consider the attacker gets detected at stage $k$ whenever the action pair \{Defend, Attack\} gets selected. The game \emph{stops} at stage $k$ if the attacker gets detected a total of $L$ times until stage $k$, known as the stopping state. An illustration of a simple graph with an edge-game and $L=1$ in shown in Figure~\ref{fig:MZSG-illustration}. This stopping state models the fact that after getting detected a total of $L$ times, the attack is permanently disabled. 
To define termination of an edge-game formally, consider the indicator function at the $k$-th stage which is given by
\[
I(i_{k},j_{k}) = \begin{cases}
1 & \text{if } \{i_{k},j_{k}\} = \{D,A\}, \\
0 & \text{otherwise}.
\end{cases}
\]
Then, the game stops if there exists an $t < K_e$ for which $ \sum_{k=1}^{t} I(i_{k},j_{k}) = L$. Now, given a sequence of player actions $\{(i_1, j_1), \dots, (i_{K_e}, j_{K_e})\}$, the net payoff to the attacker is given by
\[
J = \sum_{k=1}^{t-1} S^k_{i_k, j_k} + S_{ND,NA}, 
\]
since the game stops at stage $t \leq K_e$. The quantity $S_{ND,NA}$ indicates the cost at a stopping state defined by the mobility cost. This paper analyzes an edge-game for $L := 1$, although the approach can be extended to $L\geq 1$ as well. 

The edge-game is dynamic in nature since we use the space of behavioral policies for the players. For an edge-game a \emph{multi-stage behavioral policy} \cite{hespanha2017noncooperative} for the defender is a set of probability distributions $\mathcal{Y}_e := \{y_1, \dots, y_{K_e}\} \in \Delta_{2}^{K_e}$ and similarly for the attacker is a set of probability distributions $\mathcal{Z}_e := \{z_1, \dots, z_{K_e}\} \in \Delta_{2}^{K_e}$, where $\Delta_{2}$ is the probability simplex in $2$ dimensions. Let $J_{K_e}: \Delta_{2}^{K_e} \times \Delta_2^{K_e} \to \mathbb{R}$ denote the expected payoff to the attacker in the edge-game.It can be shown that $J_{K_e}$ is obtained from a forward recursive equation which is defined as,
\begin{equation}
\label{eq:J_e}
J_a = \sum_{k=1}^{a}y_{k}'S^kz_{k} - \sum_{b=1}^{a-1}y_{b,1}z_{b,1}J_{a-b},
\end{equation}
where $J_a$ is the expected pay-off at stage $a \in {1,2,\dots,K_e}$. 
If the attacker attacks an edge $e$, then the \emph{cost} along the edge is defined by a pair of behavioral policies $(\mathcal{Y}_e^*, \mathcal{Z}_e^*)$ that are in \emph{Nash equilibrium} \cite{hespanha2017noncooperative}, i.e., they satisfy,
\[
J_{K_e}(\mathcal{Y}_e^*, \mathcal{Z}_e) \leq J_{K_e}(\mathcal{Y}_e^*, \mathcal{Z}_e^*) \leq J_{K_e}(\mathcal{Y}_e, \mathcal{Z}_e^*), \, \forall \mathcal{Y} , \mathcal{Z} \in \Delta_{2}^{K_e}.
\]
We denote the outcome of an edge-game as $J_{K_e}^* := J_{K_e}(\mathcal{Y}_e^*, \mathcal{Z}_e^*)$. Since the cost of traversing an edge $e$ depends on whether the edge has been attacked or not, we define the cost by $w_e$ given by,
\begin{equation}
\label{eq:edge_weight}
w_e = \begin{cases} J_{K_e}^*, &\text{ if edge $e$ is attacked, } \\
\sum_{k=1}^{K_e} S^k_{ND, NA}, &\text{ otherwise,}\end{cases},
\end{equation}
where the summation denotes the mobility cost over the edge in the absence of any attack. Let $e_{i,j}$ denote the edge connecting vertices $i$ and $j$ if it exists. Let $S, T \in V$ denote a start/destination pair of vertices. A path from $S$ to $T$ is a collection of the edges $\pi_{ST} := \{e_{Si_1} ,e_{i_1 i_2}, \dots, e_{i_n T} \}$. The collection of paths from $S$ to $T$ is denoted as $P_{ST}$. The cost of a path $\pi_{ST} \in P_{ST}$ is defined as
\begin{equation}\label{eq:we}
w_{\pi_{ST}}  = \sum_{e \in \pi_{ST}} w_e.
\end{equation}
This cost $w_e$ can then be used to define a \emph{meta-game} being played between the path defender and the edge attacker in which the attacker selects a subset $\mathcal{E} \subset E$ of $\abs{\mathcal{E}}$ edges to attack, and the defender selects a path $\pi_{ST} \in P_{ST}$. The subset of attack edges is defined as, $\mathcal{E} = \cap_{i=1}^{\abs{P_{ST}}} \pi_{ST}^{i} \in P_{ST}$. In this work we restrict the attack only over an edge. The meta-game is represented equivalently through the entries of a matrix $W$ having a number of rows equal to the cardinality $\abs{P_{ST}}$ and the number of columns equal to cardinality $\abs{\mathcal{E}}$. A \emph{mixed policy} for the defender (resp.~attacker) in the meta-game is a probability distribution $\hat{y}$ (resp.~$\hat{z}$) over the set of paths $P_{ST}$ (resp.~the sets of edges $\abs{\mathcal{E}}$). The Nash equilibrium value for the meta-game defined as,
\begin{align}
\label{eq:MGZS-sol}
W_{NE} = \min_{\hat{y} \in \Delta_{\abs{P_{ST}}}}\max_{\hat{z} \in \Delta_{\abs{\mathcal{E}}}} \hat{y}^TW\hat{z}.
\end{align}
The optimal policies $\hat{y}^*$ and $\hat{z}^*$ represent the probability of picking a path $\pi_{ST}$ and attacking an edge $\mathcal{E}$, respectively. Clearly, the complexity of this solution scales undesirably (combinatorially) with the size of the roadmap. Therefore, a second goal in this paper is to design a computationally efficient approach to finding a secure path.


\section{Dynamic Game with Stopping States \\on an Edge}
\label{sec:mult-stage-game-edge}
In this section we analyze the multi-stage zero-sum game with L=1 i.e., the edge-game when the attack is disabled when the action pair \{Defend, Attack\} occur simultaneously at any stage of the game.

As illustrated in Figure~\ref{fig:MZSG-illustration}, the game stops either in the states denoted by $SS$ or at any of the other leaf nodes. Since this is a complete feedback game, we use a standard technique (e.g., see \cite{hespanha2017noncooperative}) to solve the recursive equation,
\begin{align}\label{eq:V_k - lvl1}
V_{k-1} & = \Val(V_{k}D + S),
\end{align}
where $k \in \{K_e, K_e -1, \dots, 1\},$ is the stage, $V_{k}$ is the expected value of the game at stage $k$, $S$ is the stage cost matrix, $D$ is the decision matrix at every stage as a consequence of the termination at the game corresponding to the action pair $\{\text{Defend, Attack}\}$ and $\Val(\cdot)$ is a function mapping which takes in expected value of the next stage $k$ and stage cost and returns the expected cost stage $k-1$. For the $2\times 2$ case considered in this paper, equation~\eqref{eq:V_k - lvl1} is further expanded as, 
\begin{align}\label{eq:V_k - lvl2}
V_{k-1} &= \Val \left( V_{k}\underbrace{\begin{bmatrix} 0 & 1 \\ 1 & 1 \end{bmatrix}}_{D}+ \begin{bmatrix} s_{11} & s_{12} \\ s_{21} & s_{22} \end{bmatrix} \right) \nonumber \\
&= \min_{y_k \in \Delta_2} \max_{z_k \in \Delta_2} y_{k}' \left( V_{k}\begin{bmatrix} 0 & 1 \\ 1 & 1 \end{bmatrix} + \begin{bmatrix} s_{11} & s_{12} \\ s_{21} & s_{22} \end{bmatrix} \right)z_{k}.
\end{align}
The expected value of the game for given stage is given as, 
\begin{align}\label{eq:V_k - final form}
V_{k-1} = {y_{k}^*}' \left( V_{k}\begin{bmatrix} 0 & 1 \\ 1 & 1 \end{bmatrix} + \begin{bmatrix} s_{11} & s_{12} \\ s_{21} & s_{22} \end{bmatrix} \right)z_{k}^*,
\end{align}
where $\{y_{k}^*, z_k^*\}$ is a mixed Nash equilibrium at stage $k$. 

Our first result summarizes analytic expressions for the Nash equilibrium policies and the value below.
\begin{theorem}
	\label{th:L1_pv}
	The unique Nash equilibrium and value of an edge-game (equation~\eqref{eq:V_k - final form}) at any stage $\forall k \in \{1,2,\dots,K_{e}\}$ with stopping state is given by,
	\begin{align}
	y_{k}^{*} & = \begin{bmatrix} \dfrac{s_{22} - s_{21}}{s_{11} - s_{12} - s_{21} + s_{22} - V_{k}} \\[2ex] \dfrac{s_{11} - s_{12} - V_{k}}{s_{11} - s_{12} - s_{21} + s_{22} - V_{k}}
	\end{bmatrix}, \text{ and }
	\label{eq:policy_def_L1}
	\end{align}
	\begin{align}
	z_{k}^{*} & = \begin{bmatrix} \dfrac{s_{22} - s_{12}}{s_{11} - s_{12} - s_{21} + s_{22} - V_{k}} \\[2ex] \dfrac{s_{11} - s_{21} - V_{k}}{s_{11} - s_{12} - s_{21} + s_{22} - V_{k}}
	\end{bmatrix},
	\label{eq:policy_att_L1}
	\end{align}
	where $V_{K_{e}} := 0$, and  
	\begin{align}
	V_{k-1} = V_{k} + \frac{\det{S} - s_{22}V_{k}}{s_{11} - s_{12} - s_{21} + s_{22} - V_{k}}, 
	\label{eq:V_k_recur_th1}
	\end{align}
	and $\det{S}$ is the determinant of $S$.  \frqed
\end{theorem}

\medskip

The proof of this result has been included in the appendix. Theorem~\ref{th:L1_pv} yields a formula to compute the solution of an egde-game recursively. Next, we parameterize the stage cost matrix $S$ with ratios $r_1$ and $r_2$ as follows,
\begin{equation}
\begin{split}
S & = s_{11} \begin{bmatrix} 1 & 1 \\[1ex] r_1 & r_2
\end{bmatrix},
\end{split}
\label{eq:S_mat_parameterized}
\end{equation}
\[
\text{where }r_1:= \dfrac{s_{21}}{s_{11}}, \text{ and } r_2:=\dfrac{s_{22}}{s_{11}}.
\]
The motivation for such a matrix relies on the assumption that, the payoff for defending remains independent of the actions taken by the attacker. The cost of defense is represented by $s_{11}$, whereas $s_{12}$ represents the security loss. Mobility cost is represented by $s_{22}$ i.e., the minimum cost defender pays going from the current to next stage.  Furthermore, we assume $r_1 \geq 1$ and $r_2 < 1$ for rest of the paper. The condition of $r_1 \geq 1$ naturally fit the incentive of an attacker to cause a loss. Using a parameterized matrix \eqref{eq:S_mat_parameterized} we obtain the following equation,
\begin{equation}
\label{eq:V_rec_s22_0}
V_{k-1} = s_{11}\left(V_{k} + \frac{r_2 - r_1 - r_2V_{k}}{r_2 - r_1 - V_{k}}\right).
\end{equation}
With the game well defined in the final stage $K_e$ i.e., $V_{K_e}:=0$, we can further determine the limiting values of the attacker and defender policy. The Nash equilibrium at the final stage $K_e$ is given by,
\begin{align}
\label{eq:policy_att_def_K}
y_{K_e}^{*} & = \begin{bmatrix}\text{}\\ 1\\[2ex] 0 \vspace{2ex}
\end{bmatrix},
z_{K_e}^{*}  = \begin{bmatrix} \dfrac{1 - r_2}{r_1 - r_2} \\[2ex] \dfrac{r_1 - 1}{r_1 - r_2}
\end{bmatrix}.
\end{align}

The following result summarizes a key property of this analysis.

\medskip

\begin{corollary}\label{cor:initialNE}
	Given an edge-game with a constant stage cost $S$ defined by equation~\eqref{eq:V_k - lvl1}, the mixed policy for the defender and attacker at the \emph{start} of the game in the limit of stages $K_e \to \infty$ is given as,
	\[
	\lim_{K_{e} \to \infty} y_{1}^{*}  = \begin{bmatrix} 0 \\[2ex] 1
	\end{bmatrix}, \, 
	\lim_{K_{e} \to \infty} z_{1}^{*}  = \begin{bmatrix} 0 \\[2ex] 1
	\end{bmatrix}. 
	\] 
\end{corollary}

\medskip

In short, this means that both players begin with not defending and not attacking, respectively at the start of an edge-game and gradually (monotonically) shift the weight toward defending and attacking as the stages progress.

\subsection{Analytical, approximate solution and numerical evaluation}
We determine two analytical solutions to the defined recursion equation namely; $r_2 > 0 $ and when the ratio $r_2 = 0$, while normalizing the payoff corresponding to defense to unity, i.e., $s_{11}= 1$. Furthermore, we determine an approximate solution with bounded error.

\medskip

\begin{proposition}\label{prop:lowerboundV}
	Analytical solution; In the limit of inter-stage time interval tending to zero, the value of an edge-game at any stage $k$ is given by,
	\begin{equation}
		\begin{cases}
		\text{if } r_2 = 0; \\ \quad V_k = -r_1 + \sqrt{r_1^2 + \left(2r_1K_{e} + 2r_1(1-k) + 1\right)}, \\[2ex]
		\text{otherwise, $V_k$ satisfies} \\[1ex] \quad \dfrac{(r_2 - 1)c \log(|r_2V_k - c|/r_1)}{r_2^2} -\dfrac{(V_k-1)}{r_2}  = ({K_e} - k).
		\end{cases}
	\label{eq:V_2p_cf_final}
	\end{equation}
	where $c:= r_2 - r_1$. \frqed
\end{proposition}

\medskip
\begin{proposition}\label{prop:approxsolution}
	Approximate solution; Similarly, under the same limit of inter-stage time interval tending to zero, the value of an edge-game at any stage $k$ is given by,
	\begin{equation}
	 V_k \approx -r_1 + \sqrt{r_1^2 + \left(2r_1K_{e} + 2r_1(1-k) + 1\right)} + r_2(K_e - k)
	\label{eq:V_approx}
	\end{equation}
\end{proposition}

\begin{figure*}[h]
	\begin{center}
		\subfloat[]{\includegraphics[width = 0.25\linewidth]{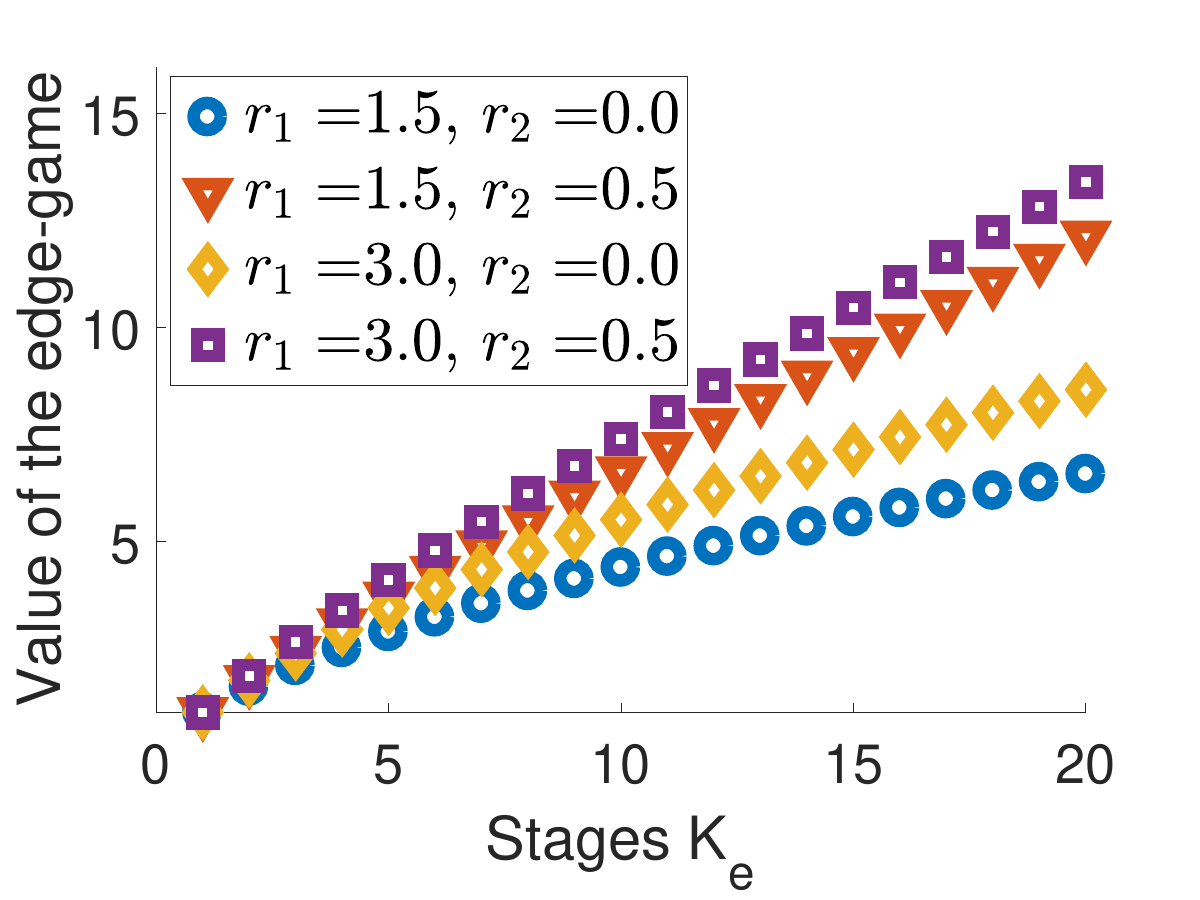}
			\label{fig:MSZG_r2_V}}
		\subfloat[]{\includegraphics[width = 0.25\linewidth]{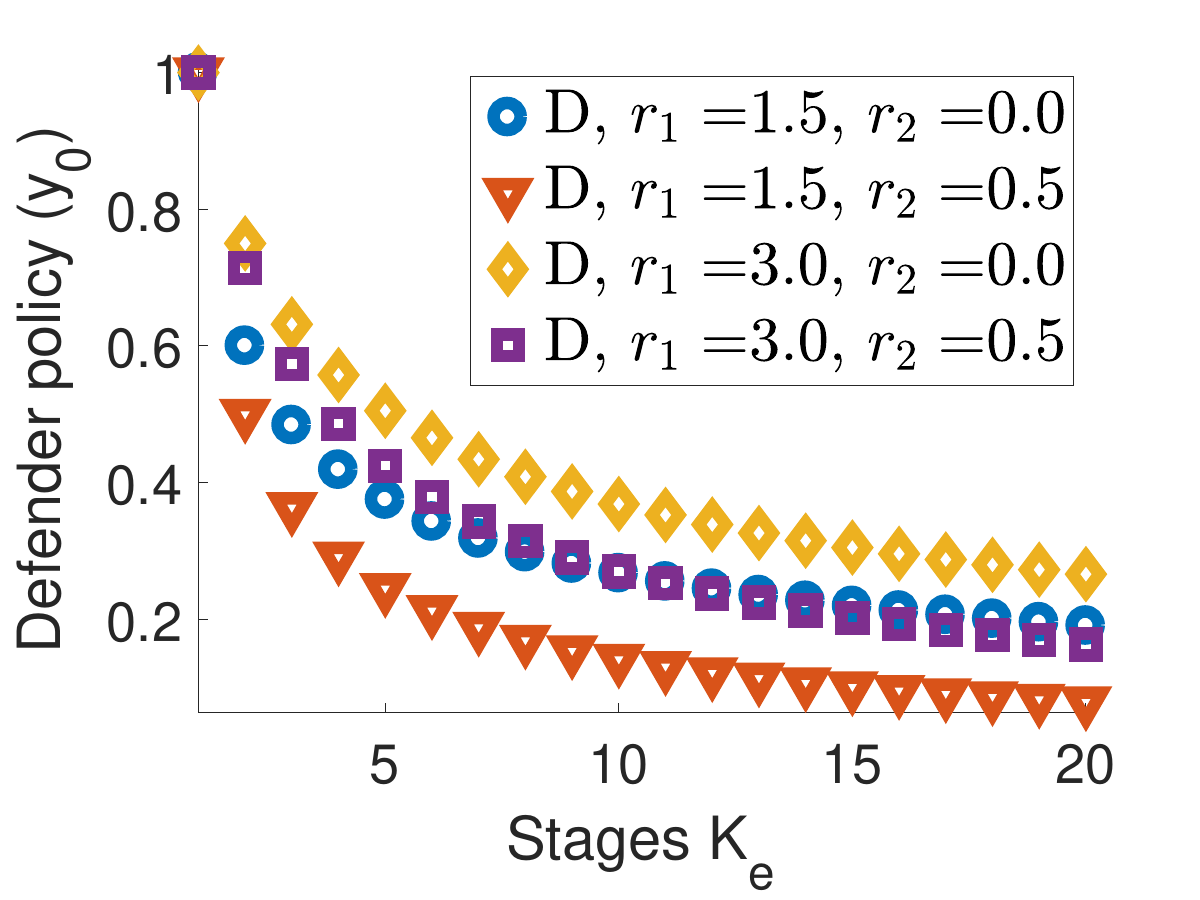}
			\label{fig:MSZG_r2_def}}
		\subfloat[]{\includegraphics[width = 0.25\linewidth]{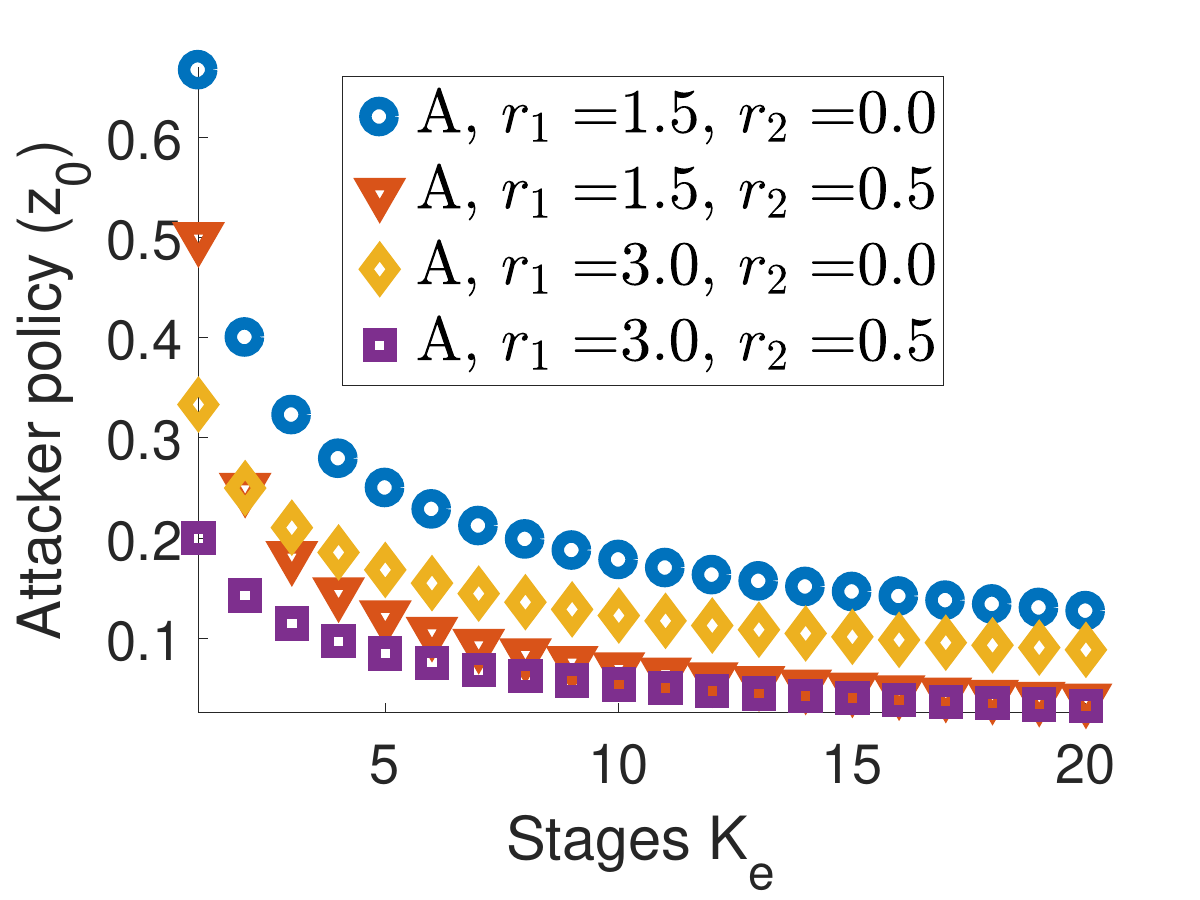}
			\label{fig:MSZG_r2_att}}
		\subfloat[]{\includegraphics[width = 0.25\linewidth]{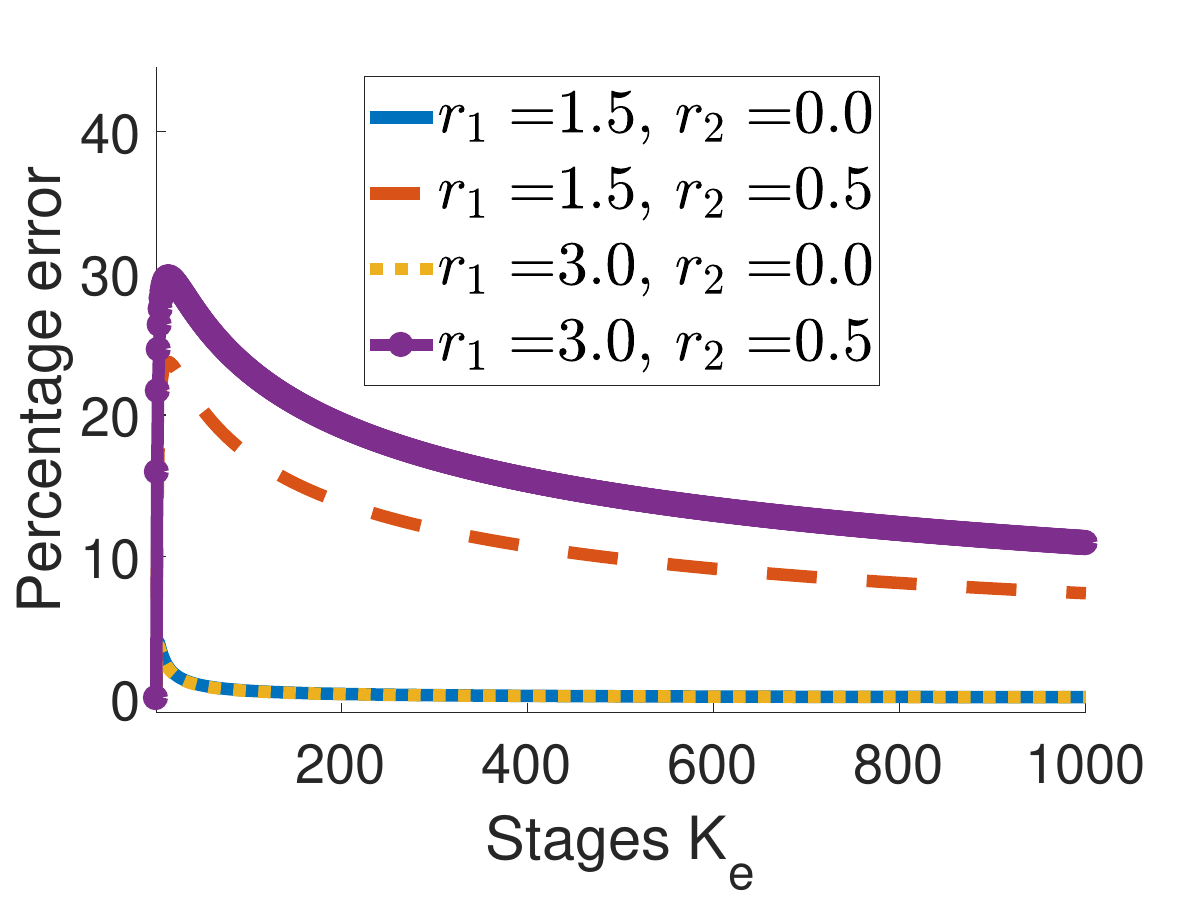}
			\label{fig:Val_error}}
		\caption{\small (a) Value of the edge-game vs. number of stages $K_e$ for a given set of $r_2$ and $r_1$ with number of attacks limited to $L = 1$. (b) Policy of the defender at stage $k=0$ vs. $K_e$ (number of stages) corresponding to the edge-game in Figure~\ref{fig:MSZG_r2_V}. (c) Policy of the attacker at stage $k=0$ vs. $K_e$ for the same conditions as in Figure~\ref{fig:MSZG_r2_att}. (d) Percentage error between approximate value (equation~\eqref{eq:V_approx}) and recursive value (equation~\eqref{eq:V_rec_s22_0}) of the game for a set of game parameters.}	
	\end{center}
	\label{fig:L_1_game}
	\vspace{-0.3in}
\end{figure*}

The proof of both the propositions are presented in the appendix.
Given the number of stages along an edge $K_e$, the value of an edge-game is determined at $k=0$, i.e., $V_0$ corresponding to the solution of game (equation~\eqref{eq:J_e}). From equation \eqref{eq:V_2p_cf_final} we can deduce that, the value of an edge-game increases sublinearly with total number of stages $K_e$ for sufficiently small values of $r_2$. The effect $r_2$ on the value of an edge-game, policy of the defender and attacker as a function of the total stages are shown in Figures~\ref{fig:MSZG_r2_V}, \ref{fig:MSZG_r2_def} and \ref{fig:MSZG_r2_att}. It is observed that with increasing $r_2$ the sublinearity property dominated by a linear term which is a function of $r_2$ also indicated by equation~\eqref{eq:V_approx}. The policy of the defender $y_0$ at the start of a game is shown in the Figure ~\ref{fig:MSZG_r2_def} and it is observed that the probability of defending reduces with the increasing number of stages $K_e$. Similarly, from Figure ~\ref{fig:MSZG_r2_att}, the attack policy $z_0$ is shown to decrease over the number of stages $K_e$. The results of the policies for both the defender and  attacker is a consequence of the additive stage cost, where it is  the most beneficial for the attacker and defender to play actively (Attack, Defend) toward the end of the game. Finally, we compare the approximate value in equation~\eqref{eq:V_approx} with the recursive value in equation~\eqref{eq:V_rec_s22_0} and plot the percentage error with respect to the recursive value in Figure~\ref{fig:Val_error}. We observe that the accuracy of the approximation is higher with decreasing $r_1$ and $r_2$, and the error for any given $r_1$ and $r_2$ tends to zero with larger number of stages $K_e$, thus the error is bounded.
In the next section, the results of the edge-game will be used to create a meta-game over the possible paths and edges.

\section{Solution to the Meta-game}
\label{sec:meta-game}
Consider any graph $G$ with vertices $V$ and edges $E$ as defined in Section \ref{sec:model} with each edge $e \in E$ being associated with a stage $K_e$ respectively. The cost $V_0$ over an edge $e$ with the number of stages, $K_e$ is determined using an edge-game presented in Section \ref{sec:mult-stage-game-edge}. The edge-game solutions are used as edge weights for the corresponding edge $e$ of the graph $G$.  Suppose that there exists $m$ possible paths from a source vertex $S$ to a destination vertex $T$ and contains a total of $n$ edges over the $m$ possible paths, the meta-game matrix is defined as, 
\[
W 
= \begin{bmatrix} 
W_{\pi_1e_1} & W_{\pi_1e_2} & \dots  & W_{\pi_1e_n}\\
W_{\pi_2e_1}  & W_{\pi_2e_2} & \dots  & W_{\pi_2e_n} \\
\dots  & \dots & \dots & \dots\\
\dots  & \dots & \dots & \dots\\
W_{\pi_me_1}  & W_{\pi_me_2} & \dots  & W_{\pi_me_n} \\ 
\end{bmatrix},
\]
where $\pi_{i}, \ i \in \{1,2,\dots,m\}$ are possible paths and $e_{j}, \ j \in \{1,2,\dots,n\}$ are possible edges. Path $\pi_{i}$ contains $p_{i}$ linked edges. $W_{\pi_ie_j}$ represents the sum of edge costs on the path $\pi_i$ given that the attack is over edge $e_j$, and is given by,
\begin{equation}
\label{eq:meta-game_W_matrix}
W_{\pi_{i}e_{j}} = \begin{cases}
\sum_{x=1}^{p_{i}} \sum_{k=1}^{K_{e_{x}}} S^{k}_{ND,NA}, & \text{if } e_{j} \notin \pi_{i} \\
\sum_{e_{j} \in \pi_{i}} w_{e_{j}}, & \text{otherwise}.
\end{cases}
\end{equation}
The if condition refers to the cost of mobility under the assumption that, the entire path is free of any attack. The later condition refers to the cost of a path when under an attack along one of its edges, defined in equation~\eqref{eq:we}. The zero-sum meta-game $W$ is solved using a standard linear programming technique \cite{hespanha2017noncooperative} to obtain a \emph{secure route(s)}. The meta-game solution $W_{NE}$ is compared against the length of the shortest path $L_{SEA}$ (shortest path heuristic) following the same constraints that, only one of the edge $e$ in the graph $G$ can be attacked and defended. $L_{SEA}$ is computed using the Algorithm~\ref{algo:SP}. Here, $\pi_{SP}$ denotes the shortest path.
\begin{algorithm}[h]
	\textbf{Input:} G(graph) \\
	\SetKwFunction{Dijkstra}{Dijkstra}
	\SetAlgoLined
	\KwOut{$L_{SEA}$ $:=$ Length of the shortest path under with an edge attack}
	\For{ every $e \in E$}{
		Set $w_e = V_{0}$ for edge $e$ \;
	} 
	$\pi_{SP}$  = \Dijkstra($V, \{w_{e_1}, \dots, w_{e_{\abs{E}}} \}$) \\
	Compute the row of $W$ corresponding to $W_{{\pi_{SP}},e}$.\\
	$L_{SEA}$ =	$\argmax_{e \in \pi_{SP}} W_{{\pi_{SP}},e}$
	\caption{Shortest path edge attack}
	\label{algo:SP}
\end{algorithm}

\begin{figure*}[h]
	\begin{center}
		\subfloat[]{\includegraphics[width = 0.24\linewidth]{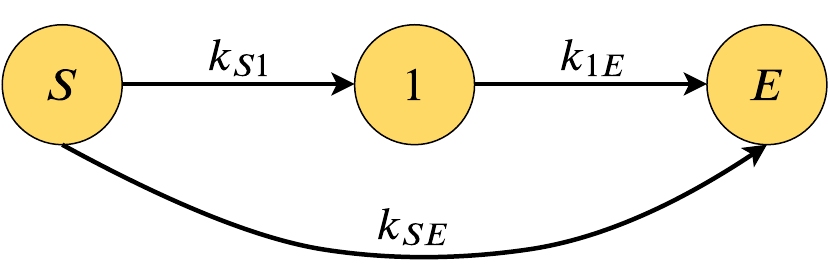}
			\label{fig:simple_network}	
		}
		\subfloat[]{\includegraphics[width = 0.18\linewidth]{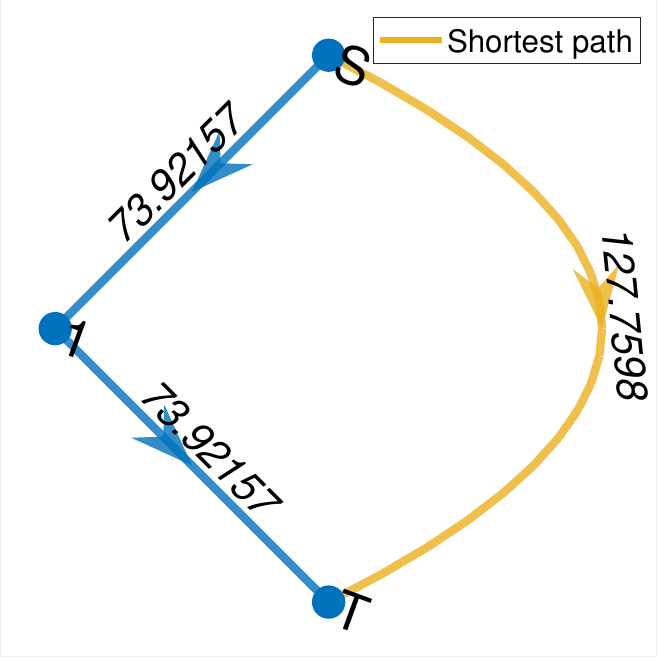}
			\label{fig:simple_network_sol}	
		}
		\subfloat[]{\includegraphics[width = 0.26\linewidth]{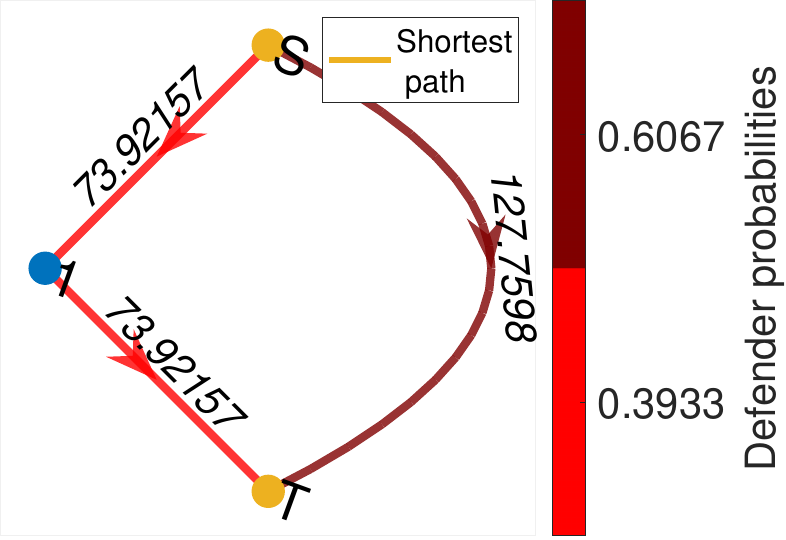}
			\label{fig:simple_network_def}	
		}
		\subfloat[]{\includegraphics[width = 0.26\linewidth]{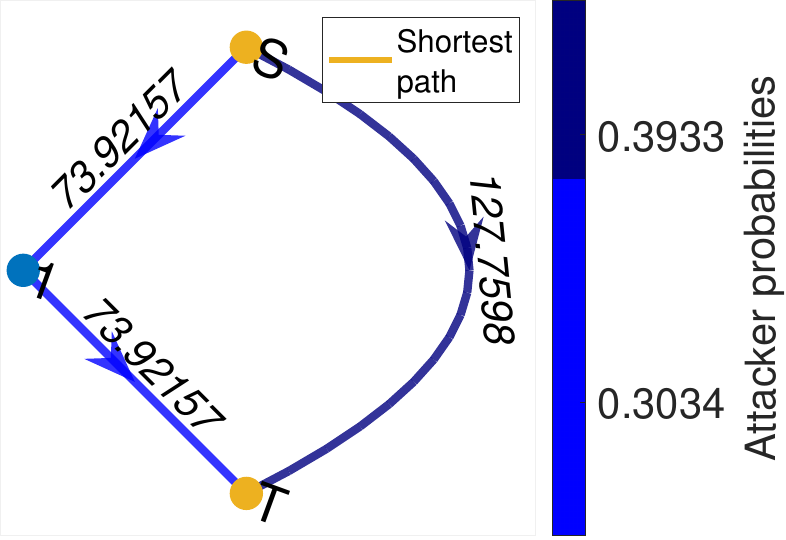}
			\label{fig:simple_network_att}	
		}
		\caption{\small (a) Illustration of a simple graph with 3 vertices and 3 edges. The start and end vertex is indicated with $S$ and $T$ respectively. The number of stages between the nodes $i$ and $j$ are given by $k_{i,j}$. (b) The simple network (figure \ref{fig:simple_network}) with stages over the edge, $k_{S1} = k_{1T} = 3$ and $k_{ST} =6$. The shortest path is calculated over the edge weights. (c) The solution of the simple graph meta-game with the defender probabilities over the paths. The shortest path is indicated with a larger arrow as compared to others and with lighter shade of vertex. (d) The solution of the simple graph meta-game with the attacker probabilities over the edges.}	
	\end{center}
	\label{fig:simple_network_game}
	\vspace{-0.33in}
\end{figure*}

An illustration of a simple graph is presented in Figure \ref{fig:simple_network}. The graph consists of 2 paths namely; $P_{ST} = \{\{e_{ST}\},\{e_{S1},e_{1T}\}\}$ i.e., from vertex $S$ to $T$, and from vertex $S \to 1$ followed by $1 \to T$. The edges consist of $\mathcal{E} = \{e_{ST}, e_{S1}, e_{1T}\}$. We assume a fixed stage cost given by,
\[
S = \begin{bmatrix}
30 & 30 \\ 70 & 10
\end{bmatrix},
\]
over the entire graph $G$, and solve for the edge-game and meta-game. Results of the simple graph are illustrated in Figure \ref{fig:simple_network_sol}, also depicting the shortest path which resulted from Algorithm~\ref{algo:SP}. The defender and attacker probabilities shown in Figures \ref{fig:simple_network_def} and \ref{fig:simple_network_att}, respectively. From the plots, we observe that the probability of choosing the shortest path by the defender and an edge on the same by the attacker is higher as compared to the alternative. We observe that, the results are dependent on the stages $K_e$ and stage costs along each edge $e$. 

\begin{figure*}[h]
	\begin{center}
		\subfloat[]{\includegraphics[width = 0.25\linewidth]{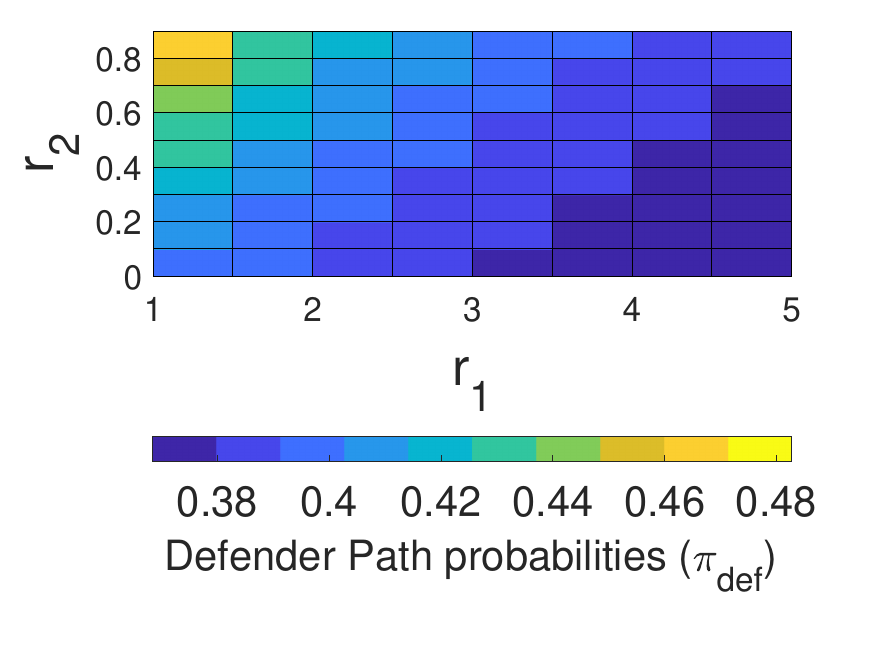}
			\label{fig:simple_network_sen_def_r1r2}	
		}
		\subfloat[]{\includegraphics[width = 0.25\linewidth]{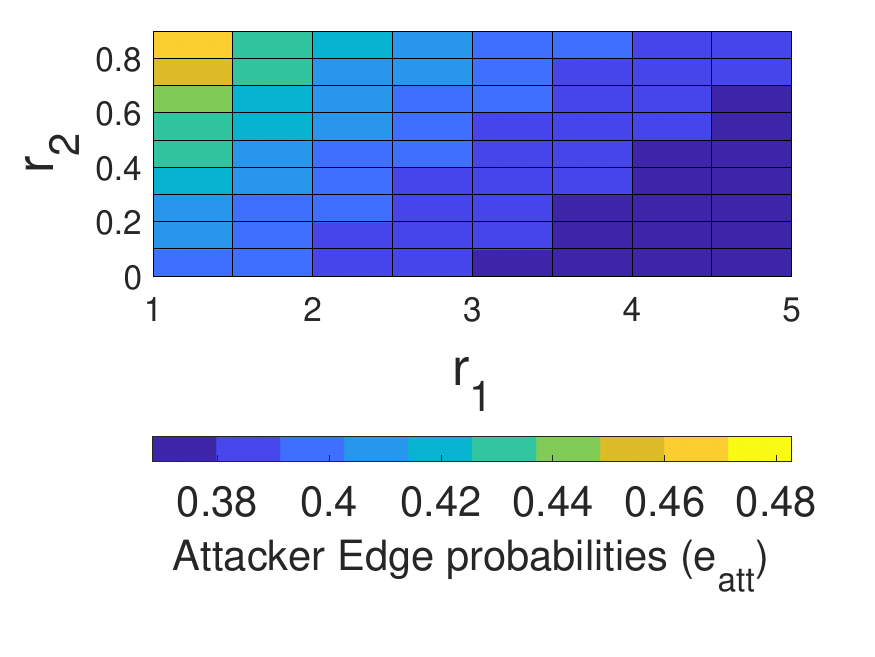}
			\label{fig:simple_network_sen_att_r1r2}	
		}
		\subfloat[]{\includegraphics[width = 0.25\linewidth]{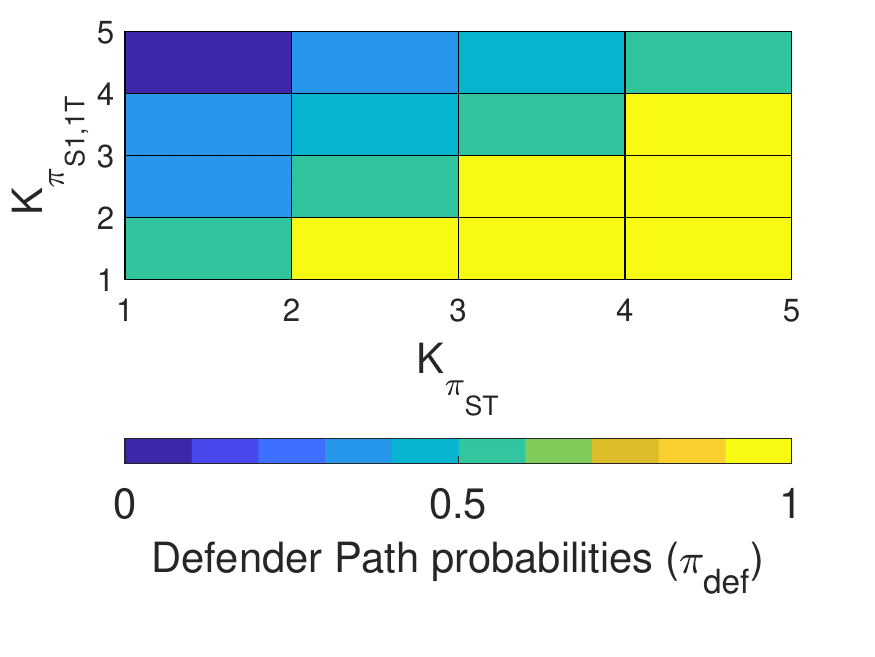}
			\label{fig:simple_network_sen_path}	
		}
		\subfloat[]{\includegraphics[width = 0.25\linewidth]{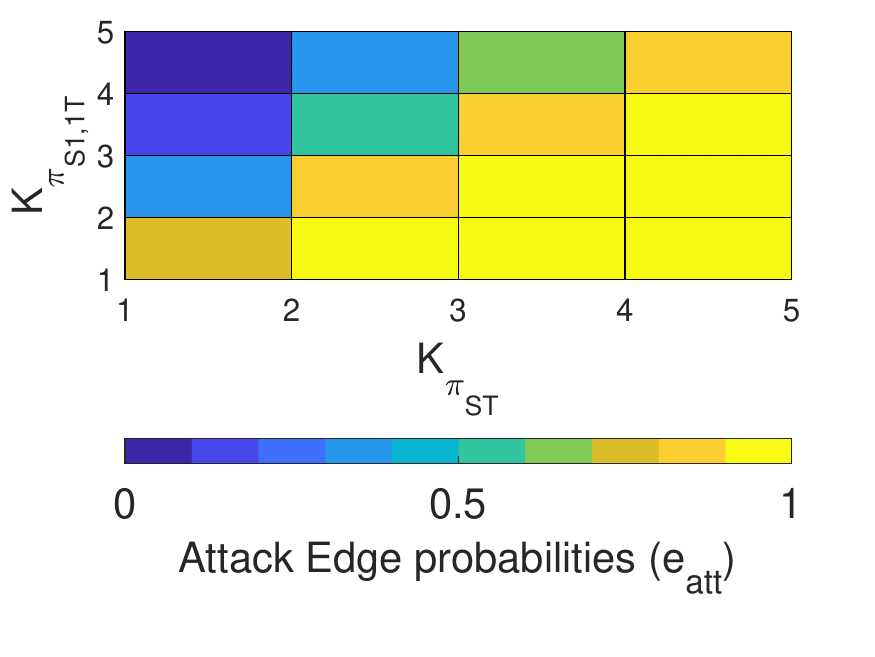}
			\label{fig:simple_network_sen_edge}	
		}
		\caption{\small (a) The sensitivity of choosing the shortest path ($\pi_{\text{def}}$) with changing $r_1$ and $r_2$ with fixed stages over each edge. (b) The sensitivity of choosing the shortest path edge ($e_{\text{att}}$) with changing $r_1$ and $r_2$ with fixed stages over each edge (c) The sensitivity of choosing the shortest path ($\pi_{\text{def}}$) with changing number of stages over the edges $K_{\pi_{S1,1T}}$ and $K_{\pi_{ST}}$ given a fixed stage cost. (d) The sensitivity of choosing the shortest path edge ($e_{\text{att}}$) with changing number of stages over the edges $K_{\pi_{S1,1T}}$ and $K_{\pi_{ST}}$ given a fixed stage cost}.	
	\end{center}
	\label{fig:simple_network_sensitivity}
	\vspace{-0.33in}
\end{figure*}

\subsection{Sensitivity of optimal policies to the game parameters}
Here we observe the sensitivity of defender (paths) and attacker (edges) policies for the simple graph $G$ (Figure ~\ref{fig:simple_network}) as a function of stage cost entries and stages $K_e$ along an edge $e$. In the first scenario, we examine the sensitivity over stage costs. The stage cost is parameterized with 2 ratios also defined in equation \eqref{eq:S_mat_parameterized} by,
\[
S = s_{11}\begin{bmatrix}
1 & 1 \\ r_{1} & r_{2}
\end{bmatrix},
\] 
We abbreviate the shortest path as $\pi_{\text{def}}$ and edge along the same as $e_{\text{att}}$. The sensitivity plot for both $\pi_{\text{def}}$ and $e_{\text{att}}$ for changing $r_1$ and $r_2$ are shown in Figures~\ref{fig:simple_network_sen_def_r1r2} and \ref{fig:simple_network_sen_att_r1r2}, respectively. The probability of picking the shortest path and the corresponding edge decreases with increasing $r_1$, indicative of the fact that, the defender (resp. attacker) is aware of risks and chooses to avoid the shortest path. For increasing $r_2$, we observe that the probability of choosing the shortest path increases. This is related to the cost of mobility, i.e., under no defense and no attack the payoff is high, therefore the defender (resp. attacker) wants to reach the destination with minimum number of stages.

\medskip

Next, we characterize the sensitivity of choosing $\pi_{\text{def}}$ and $e_{\text{att}}$ with varying number of stages $K_e$ along an edge $e$ i.e., along shortest path and the alternate path which consists of 2 edges. We increase the number of stages on both edges equally. From Figure \ref{fig:simple_network_sen_path}, it can be inferred that, the probability of choosing the shortest path $\pi_{\text{def}}$ (resp. alternate path) monotonically increases (resp. decreases) with the number of stages. Similarly, from Figure~\ref{fig:simple_network_sen_edge}, the probability of choosing the shortest path edge $e_{\text{att}}$ is directly proportional to the number of stages over the edge $e_{ST}$ and is inversely proportional to the number of stages along the path $e_{S1}\cup e_{1T}$. Thus, if the number of stages over a path is significantly higher than over other paths, then the defender probability of selecting such a path is higher.

\medskip

We conclude from the sensitivity analysis that the ratios $r_1$ and $r_2$ govern the defender propensity to be either \emph{risk seeking} or \emph{risk averse}, i.e., when the costs of mobility is high and security loss is low, the defender is more likely to choose the shortest path, i.e., the defender is risk seeking otherwise it is risk averse. Furthermore, the influence of edge stages $K_e$ strongly governs the defender and attacker policy. Alternate path(s) with multiple edges consisting of lower number of stages strongly deviate the policy away from the shortest path. In the next subsection, we will see how well the solution of meta-game performs against Algorithm~\ref{algo:SP} over larger sized graphs and determine if the shortest path obtained from Algorithm~\ref{algo:SP} can serve as a reasonable secure route for large graphs.

\section{Comparisons on larger graphs}
\label{sec:simulation}
\begin{figure*}[h]
	\begin{center}
		\subfloat[]{\includegraphics[width = 0.25\linewidth]{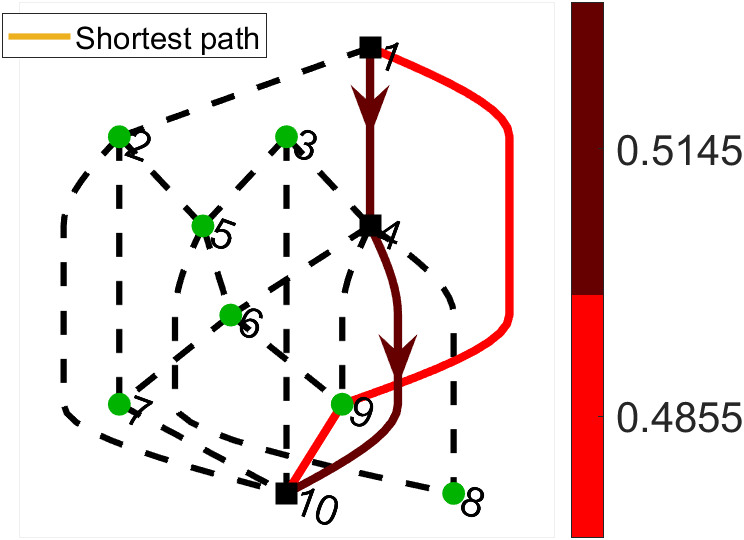}
			\label{fig:large_network_path}	
		}
		\subfloat[]{\includegraphics[width = 0.25\linewidth]{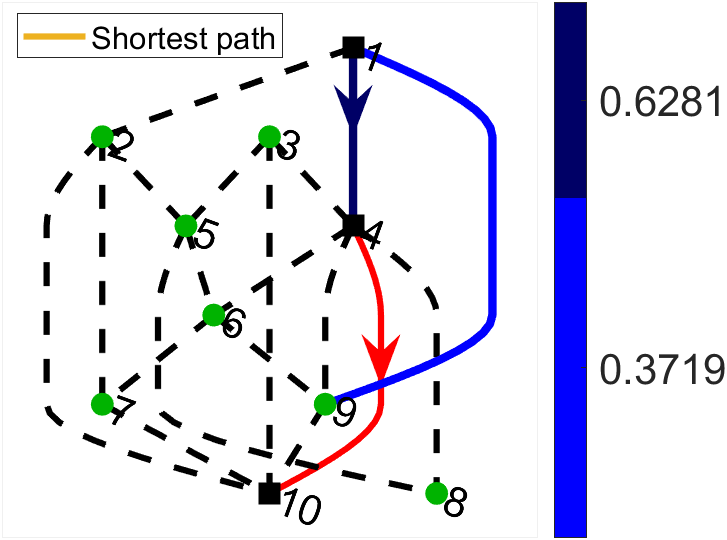}
			\label{fig:large_network_edge}	
		}
		\subfloat[]{\includegraphics[width = 0.25\linewidth]{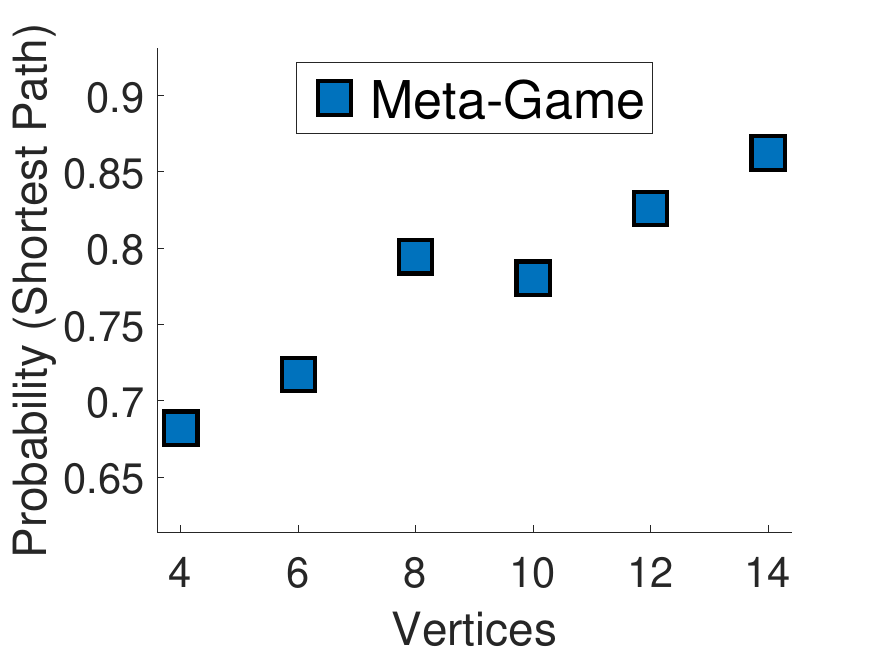}
			\label{fig:large_network_pdef}	
		}
		\subfloat[]{\includegraphics[width = 0.25\linewidth]{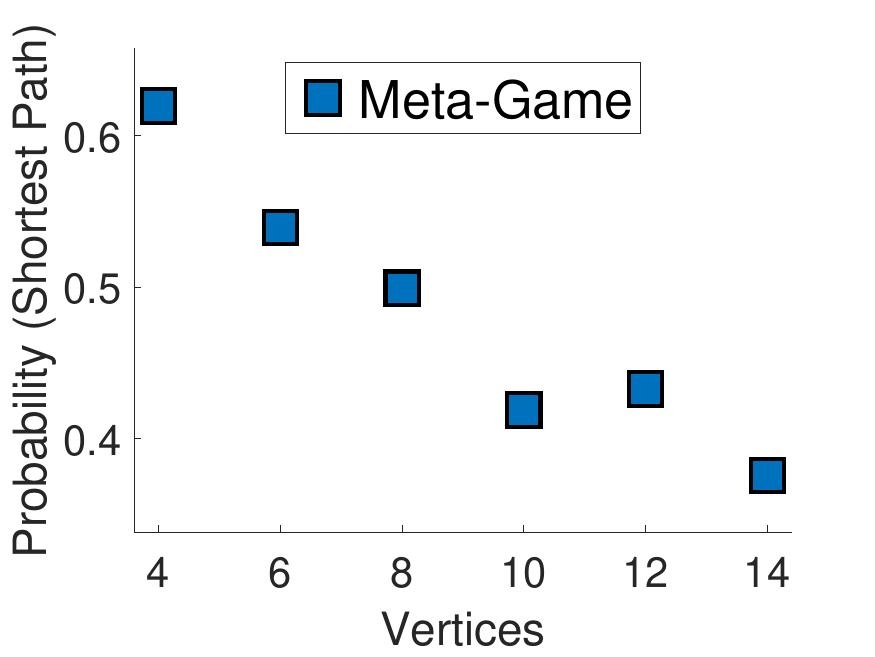}
			\label{fig:large_network_pdef_FC}	
		}
		\caption{\small (a) A graph consisting of 10 nodes which is sparsely connected. The output of Algorithm~\ref{algo:SP} is path $1-10$. Of all paths available, the path $1-4-10$ has highest likelihood of getting selected. (b) Edge $1-10$ has the least chance of being attacked, while edge $1-4$ has the highest chance of getting attacked. (c) The probability of choosing the shortest path for graphs with an average vertex degree in the interval $[2, 3]$. (d) Probability of choosing the shortest path in a fully connected network.}	
	\end{center}
	\label{fig:large_network_vs_sp}
\end{figure*}

In this section, we solve the meta-game (Section~\ref{sec:meta-game}) played over graphs of varying size to determine a secure route and compare the result against solution of Algorithm~\ref{algo:SP} treated as a baseline. The shortest path along with the probabilities of choosing the paths and edges on a sparsely connected graph with 10 vertices are shown in Figure~\ref{fig:large_network_path}  and~\ref{fig:large_network_edge}. The shortest path is indicated by square block vertex and with an arrow. The paths(s) of interest is from a source vertex $1$ to the destination vertex $10$. We observe a higher probability of picking an alternate path as opposed to a shortest path. However, the probability of choosing an attack edge is distributed across multiple paths. The results illustrate that even for a sparse graph, a \emph{secure route} is not necessarily the shortest path. 

\medskip

In general, for densely connected directed acyclic graphs (DAG) with $N$ vertices, the possible paths scale as $2^{N - 2}$ with total number of edges being $N(N+1)/2 -N$. Therefore, the size of a meta-game increases exponentially with the number of vertices leading to a meta-game matrix, $W \in \R^{\left(2^{N-2}\right) \times \left(\frac{N(N+1)}{2} -N\right)}$. Now we will investigate the solutions of the meta-game from equation~\eqref{eq:MGZS-sol} in comparison with solution of Algorithm~\ref{algo:SP} for a given graph. The connectivity of a graph is characterized by the degree of each vertex.
We observe the behavior of the meta-game on sparsely and fully connected graphs. For a sparse graph the degree of each vertex is less than the number of nodes (assuming no self-loops). A sparse graph is generated by uniformly sampling $N$ vertices from a unit square and randomly connecting them such that a desired degree for each vertex is obtained. The number of stages $K_e$ over an edge $e$ is proportional to the Euclidean distance between the connected vertices. Finally, all the stage costs along any edge of the graph are set to a constant value (defined in Section~\ref{sec:mult-stage-game-edge}). The computation times and the solutions of both the meta-game and Algorithm~\ref{algo:SP} for sparse and fully connected graphs are compared against each other in Tables~\ref{tab:MG_L_SEA_table} and~\ref{tab:MG_L_SEA_fullC} respectively. The average degree of each vertex in sparse the graph is set between the limits $[2, 3]$. 

\medskip

From Table~\ref{tab:MG_L_SEA_table}, we observe the ratio of average time taken to solve the meta-game to Algorithm~\ref{algo:SP} decreases with an increasing number of nodes, but, with a decrease in benefit in terms of cost optimality of Algorithm~\ref{algo:SP}. The decrease in computation time is a consequence of the average degree per vertex across graphs of various sizes, thus, increasing the sparsity of graphs with large number of vertices. In contrast, from Table~\ref{tab:MG_L_SEA_fullC}, we observe in dense graphs the ratio of cost performance between the two approaches decreases with the graph size, but at the expense of increasing the ratio of computation times. Probabilities of picking the shortest path is shown in Figures~\ref{fig:large_network_pdef} and~\ref{fig:large_network_pdef_FC}. From both figures, we observe that the probabilities of picking the shortest path corresponding to Algorithm~\ref{algo:SP} increases with the sparsity of a graph as opposed to a densely connected graph. This implies that the defender becomes risk seeking over sparse graphs and risk averse over a densely connected graphs.  

\begin{table}[h]
	\centering
	\caption{\small Performance of meta-game vs. Algorithm~\ref{algo:SP} averaged over 100 runs with an average degree of 2 on each vertex.}
	\resizebox{1\linewidth}{!}{
		\begin{tabular}{cccc}
			\toprule
			\begin{tabular}[c]{@{}c@{}}Vertices\end{tabular} & \begin{tabular}[c]{@{}c@{}}Time performance, \\ $\text{Time}(W_{NE})/\text{Time}(L_{SEA})$ \end{tabular} &  \begin{tabular}[c]{@{}c@{}} Cost performance, \\ $W_{NE}/L_{SEA}$ \end{tabular} \\
			
			\midrule
			4 &           122.125 &                  0.88 \\
			6 &           122.750 &                  0.89 \\
			8 &           127.625 &                  0.91 \\
			10 &           143.250 &                  0.92 \\
			12 &           133.286 &                  0.94 \\
			14 &           105.143 &                  0.96 \\
			\bottomrule
		\end{tabular}
	}
	\label{tab:MG_L_SEA_table}
	\vspace{-0.1in}
\end{table}

\begin{table}[h]
	\centering
	\caption{\small Performance of the meta-game vs. Algorithm~\ref{algo:SP} averaged over 100 runs in a fully connected network.}
	\resizebox{1\linewidth}{!}{
		\begin{tabular}{cccc}
			\toprule
			\begin{tabular}[c]{@{}c@{}}Vertices\end{tabular} & \begin{tabular}[c]{@{}c@{}}Time performance, \\ $\text{Time}(W_{NE})/\text{Time}(L_{SEA})$ \end{tabular} &  \begin{tabular}[c]{@{}c@{}} Cost performance, \\ $W_{NE}/L_{SEA}$ \end{tabular} \\
			
			\midrule
			4 &           145.143 &                  0.87 \\
			6 &           173.571 &                  0.84 \\
			8 &           175.143 &                  0.84 \\
			10 &            99.077 &                  0.84 \\
			12 &           257.143 &                  0.85 \\
			14 &           695.900 &                  0.84 \\
			\bottomrule
		\end{tabular}
	}
	\label{tab:MG_L_SEA_fullC}
	\vspace{-0.1in}
\end{table}
\section{Conclusion and Future Directions} \label{sec:conclusion}
In this paper, we consider the case of a typical path planning problem defined over a roadmap where the vehicle searches to find an optimal path from a given source to a destination in presence of an attacker who has the ability to launch an attack over an edge of the road-map. The defender (vehicle) can take an action to detect an attack at the expense of some cost (energy), and disable the attack permanently if detected repeatedly. We modeled this setup using the framework of a zero-sum dynamic game with a stopping state being played simultaneously by the attacker and defender defined as \emph{edge-game}. We characterized the Nash equilibria of this game and provided the closed form expressions for the case where both the defender and attacker are limited to only two actions. We further provided an analytic and approximate expression for an edge-game and characterized the conditions under which it grows sub-linearly with the number of stages. The sensitivity of an edge-game was studied with respect to the (i) cost of using the countermeasure, (ii) cost of motion and (iii) benefit of disabling the attack was. The results of an edge-game was used to create a zero-sum meta-game over the roadmap. The meta-game solution was compared against the result of a well-known approach based on replacing the edge weights with the expected cost of playing the game under a single edge attack, and determining the shortest path. 

\medskip

Future directions include the extensions to multiple attacks in the edge-game i.e., $L > 1$. Non-zero sum formulations modeling different objectives of the attacker and defender is also a promising direction. The constraint of one edge attack over the graph could be relaxed to have attacks over multiple edges. Formulations involving multiple vehicles are also a topic of future investigation.

\bibliographystyle{IEEEtran}
\bibliography{references}
\appendix
\subsection{Proof of Theorem \ref{th:L1_pv}}
\noindent Since the game consists of only 2 players, the following method can be used to determine the policy for each player at any stage $k$. 
Equation ~\eqref{eq:V_k - final form} can be re-written as,
\begin{align}\label{eq:V_k - alternate form}
V_{k-1} = y_{k}' \left(\underbrace{\begin{bmatrix} a_{11} & a_{12} \\ a_{21} & a_{22} \end{bmatrix}}_{A(k)} \right)z_{k}.
\end{align}
\subsubsection*{Policy for Defender}
The expected value of a game at any stage $k$ is given as,
\begin{equation}
\label{eq:policy_p1_step1}
\begin{split}
{V}_{k}(A(k)) & \doteq \min_{y \in \mathcal{Y}} \max_{z \in \mathcal{Z}} \ y'Az, \\
& = \min_{y \in \{y_{1}, y_{2}\}} \max_{z \in \{z_{1}, z_{2}\}} \ \begin{pmatrix} (y_{1}a_{11} + y_{2}a_{21})z_{1} + \\ (y_{1}a_{12} + y_{2}a_{22} )z_{2} \end{pmatrix},
\end{split}
\end{equation}
where, $\mathcal{Y}$ and $\mathcal{Z}$ is probability simplex of dimension 2.
From \cite{hespanha2017noncooperative} the mixed policy of a player is determined through a graphical approach or analytically. Given the policy of other player(s), the expected outcome over any action must be the equal to each other. From the probability simplex of dimension 2, we get $y_{2} = 1 - y_{1}$. This leads to the following,
\begin{equation} \label{eq:policy_p1_step2}
\begin{split}
{V}_{k}(A(k)) & = \min_{y _1\in [0,1]} \max   \begin{bmatrix} y_{1} a_{11} + (1 - y_{1}) a_{21} \\ y_{1} a_{12} + (1 - y_{1})a_{22}  \end{bmatrix}.
\end{split}
\end{equation}
Equation \eqref{eq:policy_p1_step2} yields the policy for player 1 as,
\begin{equation*} \label{eq:policy_p1_final_y1}
\begin{split}
y_{1} a_{11} + a_{21} - a_{21} y_{1} & = y_{1} a_{12} + a_{22} - a_{22}y_{1}, \\
\Rightarrow y_{1}^* & =  \frac{a_{22} - a_{21}}{(a_{11} - a_{21} - a_{12} + a_{22})}. 
\end{split}
\end{equation*}
The probability of choosing the second action is given as,
\begin{equation*} \label{eq:policy_p1_final_y2}
\begin{split}
y_{2}^* & =  \frac{a_{11} - a_{12}}{(a_{11} - a_{21} - a_{12} +a_{22})}. \\
\end{split}
\end{equation*}
Substituting the values from the matrix $A$, equation~\eqref{eq:policy_def_L1} yields the optimal policy for defender.

\subsubsection*{Policy for Attacker}
Similarly, the mixed policy for attacker is determined as,
\begin{equation} \label{eq:policy_p2_step2}
\begin{split}
{V}_{k}(A(k)) & = \min \max_{z \in \{ z_{1} \}} \  \begin{pmatrix} a_{11} z_{1} + a_{12} (1-z_{1}), \\  a_{21}z_{1} + a_{22}(1-z_{1})  \end{pmatrix}. 
\end{split}
\end{equation}
Equation \eqref{eq:policy_p2_step2} gives us the policy for attacker as,
\begin{equation*} \label{eq:policy_p2_final_z1}
\begin{split}
z_{1}   a_{11} + a_{12} - a_{12}   z_{1} & = z_{1}   a_{21} + a_{22} - a_{22}   z_{1}, \\
z_{1} & =  \frac{a_{22} - a_{12}}{(a_{11} - a_{21} - a_{12} +a_{22})}. \\
\end{split}
\end{equation*}
The complimentary probability is given as,
\begin{equation*}
\label{eq:policy_p2_final_z2}
z_{2} = 1-z_1 =   \dfrac{a_{11} - a_{21}}{(a_{11} - a_{21} - a_{12} + a_{22})} .
\end{equation*}
This yields the mixed policy of the attacker in equation~\eqref{eq:policy_att_L1}.

\subsubsection*{Value of the game}
The value of the game at any stage '$k$' is given by,
\begin{equation*}
\begin{split}
V_{k-1} & = y_{k}'^{*}A(k)z_{k}^{*}, \\
V_{k-1} &= \dfrac{\det{A(k)}}{(s_{11} - s_{21} - s_{12} + s_{22} - V_{k})}. 
\label{eq:V_sol_closedform}
\end{split}
\end{equation*}
Expanding the terms,
\begin{equation}
V_{k-1} = V_{k} + \frac{\det{S} - s_{22}V_{k}}{(s_{11} - s_{21} - s_{12} + s_{22} - V_{k})}. \\
\label{eq:V_k_recur}
\end{equation}

\subsection{Proof of Proposition~\ref{prop:lowerboundV}}
To determine the lower bound of the expected outcome of the game at any stage $k$, we begin with the analysis of the recursive equation \eqref{eq:V_rec_s22_0}. It is observed that the recursive equation can be formulated by a continuous version using Taylor series expansion about the point, $k$ as follows,
\begin{equation*}
\begin{split}
V(k - \Delta k) & = V(k) - \Delta k V'(k), \\
\Rightarrow \dfrac{V(k) - V(k - \Delta k)}{\Delta k} & = V'(k), \\
\Rightarrow \lim_{\Delta k \to 0} \frac{V(k) - V(k - \Delta k)}{\Delta k} & = \dfrac{r_2 - r_1 - r_2V_k}{V_{k} - r_2 + r_1}.
\end{split}
\end{equation*}
We obtain the continuous form of equation (\ref{eq:V_rec_s22_0}) as, 
\begin{equation}
\frac{dV}{dk} = \dfrac{r_2 - r_1 - r_2V_k}{V_{k} - r_2 + r_1}.
\label{eq:V_k_cont_form}
\end{equation}
If $r_2 = 0$, integrating with respect to $V$ and $k$,
\begin{equation}
\begin{split}
\int_V^{V_{K_e}}{V_{k} + r_1} \ dV & = -\int_{k}^{K_{e}} r_1 \ ds, \\ 
\frac{V^2}{2} + r_1V |_{V}^{V_{K_e}} & = -r_1(K_e - k), \\
\frac{V_{K_e}^2}{2} + r_1 V_{K_e} - \frac{V^2}{2} -r_1V & = -r_1({K_e} - k).
\end{split}
\label{eq:V_k_cf_int}
\end{equation}
The expected value of the game at the final stage $K_{e}$ is given as $V_{K_{e}}$, which has been normalized to $1$. Substituting the value in the equation \eqref{eq:V_k_cf_int}, 
\begin{equation}
\begin{split}
\dfrac{1}{2} +r_1 - \frac{V^2}{2} - r_1V & = -r_1(K_{e} - k) .\\
\end{split}
\label{eq:V_k_cf_eq}
\end{equation}
The solution of the equation (\ref{eq:V_k_cf_eq}) yields the desired result given by,
\[
V_k  = -r_1 + \sqrt{r_1^2 + \left(2r_1K_{e} + 2r_1(1-k) + 1\right)}. 
\]
Similarly, if $r_2 > 0$, and let $c = r_2 - r_1$. Integrating the expression (\ref{eq:V_k_cont_form}) again with respect to $V$ and $k$ we obtain the implicit form of the value function as,
\begin{equation}
\begin{split}
\int_V^{V_{K_e}}\frac{V_{k} - c}{c - r_2V_k} \ dV & = \int_{k}^{K_{e}} \ ds, \\[2ex]
\dfrac{(r_2 - 1)c \log(|r_2V - c|)}{r_2^2} -\dfrac{V}{r_2} + \mathcal{K} & = ({K_e} - k),
\end{split}
\label{eq:V_k_cf_r2}
\end{equation}
where $\mathcal{K}$ is the integration constant given by
\[
\mathcal{K} := \frac{1}{r_2} - \dfrac{(r_2 - 1)(r_2-r_1) \log(r_1)}{r_2^2}. 
\]

\subsection{Proof of Proposition~\ref{prop:approxsolution}}
To determine an approximate solution of an edge-game at any stage $k$, we begin with the analysis of recursive equation \eqref{eq:V_rec_s22_0}. It is observed that the recursive equation can be divided into 2 parts namely; when $r_2 = 0$ and when $r_2 > 0$ and when the number of stages $K_e$ is large. Using the parametric recursive equation with unit defense cost ($s_{11} = 1$),
\begin{equation}
\label{eq:V_form_approx}
\begin{split}
V_{k-1} & = V_{k} + \frac{r_2 - r_1}{r_2 - r_1 - V_{k}} + \frac{- r_2V_{k}}{r_2 - r_1 - V_{k}}, \\
V_{k-1} - V_{k} & = \frac{r_2 - r_1}{r_2 - r_1 - V_{k}} + \frac{- r_2V_{k}}{r_2 - r_1 - V_{k}}.
\end{split}
\end{equation}
For a given $K_e$ the value $V_k$ at stage $k$ monotonically increases. Therefore, for a large $K_e$ as $k \to 0$, equation~\eqref{eq:V_form_approx} can be approximated as,
\begin{equation}
\begin{split}
	V_{k-1} - V_{k} & \approx \frac{- r_2V_{k}}{r_2 - r_1 - V_{k}}, \\
	V_{k-1} - V_{k} & \approx r_2
\end{split}
\end{equation}
Using the Taylor series expansion method as described in \ref{prop:lowerboundV}, we obtain the following solution,
\begin{equation}\label{eq:V_approx_r2}
V_{k-1} \approx r_2(K_e - k)
\end{equation}
Therefore, combining the solutions when $r_2 = 0$ given by equation~\eqref{eq:V_k_cf_eq} and equation~\eqref{eq:V_approx_r2} we obtain the following approximation,
\[V_{k-1} \approx -r_1 + \sqrt{r_1^2 + \left(2r_1K_{e} + 2r_1(1-k) + 1\right)} + r_2(K_e - k) \]

\end{document}